\newcommand{\Tr}[2]{\mathrm{Tr}_{#1}\left[ {#2} \right]}
\newcommand{\ket}[1]{\vert #1\rangle}
\newcommand{\bra}[1]{\langle #1\vert}

\documentclass[twocolumn,pra,superscriptaddress]{revtex4-1}
\usepackage{times}
\usepackage{booktabs}
\usepackage{bm,natbib}
\usepackage{inputenc}
\usepackage{graphicx}
\usepackage{mathrsfs}
\usepackage{dcolumn,fancyhdr}
\usepackage{amsmath}
\usepackage{amssymb}
\usepackage{amsfonts,gensymb}
\usepackage{indentfirst}
\usepackage{bbold}
\usepackage{multirow}
\usepackage{dsfont}
\usepackage{color}      

\begin{document}
\title{Unconditional preparation of nonclassical states via linear-and-quadratic optomechanics}
\author{Matteo Brunelli}\thanks{These authors contributed equally to this work}
\affiliation{Cavendish Laboratory, University of Cambridge, Cambridge CB3 0HE, United Kingdom}
\author{Oussama Houhou}\thanks{These authors contributed equally to this work}
\affiliation{Centre for Theoretical Atomic, Molecular, and Optical Physics, School of Mathematics and Physics, Queen's 
University, Belfast BT7 1NN, United Kingdom}
\affiliation{Laboratory of Physics of Experimental Techniques and Applications, University of Medea, Medea 26000, Algeria}
\author{Darren W. Moore}
\affiliation{Department of Optics, Palack\'{y} University, 17. listopadu 1192/12, 771 46 Olomouc, Czech Republic}
\affiliation{Centre for Theoretical Atomic, Molecular, and Optical Physics, School of Mathematics and Physics, Queen's 
University, Belfast BT7 1NN, United Kingdom}
\author{Andreas Nunnenkamp}
\affiliation{Cavendish Laboratory, University of Cambridge, Cambridge CB3 0HE, United Kingdom}
\author{Mauro Paternostro}
\affiliation{Centre for Theoretical Atomic, Molecular, and Optical Physics, School of Mathematics and Physics, Queen's 
University, Belfast BT7 1NN, United Kingdom}
\author{Alessandro Ferraro}
\affiliation{Centre for Theoretical Atomic, Molecular, and Optical Physics, School of Mathematics and Physics, Queen's 
University, Belfast BT7 1NN, United Kingdom}

\begin{abstract}
Reservoir engineering enables the robust and unconditional preparation of pure quantum states in noisy environments. 
We show how a new family of quantum states of a mechanical oscillator can be  stabilized in a cavity 
that is parametrically coupled to both the mechanical displacement and the displacement squared.
The cavity is driven with three tones, on the red sideband, on the cavity resonance and on the second 
blue sideband. The states so stabilized are (squeezed and displaced) superpositions of a finite number 
of phonons. They show the unique feature of encompassing two prototypes of nonclassicality for bosonic 
systems: by adjusting the strength of the drives, one can in fact move from a single-phonon- to a 
Schr\"odinger-cat-like state. The scheme is deterministic, supersedes the need for measurement-and-feedback 
loops and does not require initialization of the oscillator to the ground state. 
\end{abstract}

\maketitle

\par
\section{Introduction}
The preparation and manipulation of pure quantum states usually requires isolation of the 
system from the surrounding environment and control of the Hamiltonian. Pursuing a radically different approach, 
reservoir engineering aims instead to stabilize  genuine quantum features of a system by tailoring the properties  
of the environment~\cite{ResEng1}. Such a technique has proven particularly successful in cavity systems, where 
a damped cavity mode naturally provides a highly tunable reservoir. Reservoir engineering has been successfully 
applied to trapped atoms~\cite{ResEngAtoms} and ions~\cite{ResEngIons1,ResEngIons2,ResEngIons3}, circuit 
quantum electrodynamics~\cite{ResEngcQED1,ResEngcQED2} and opto/electro-mechanics~\cite{Kronwald,Wang,Woolley,
JieLi,Oussama}. 
Focusing  on cavity optomechanics, the stabilization of mechanical single- and two-mode squeezed states has been 
recently achieved~\cite{MechSqueezing1,MechSqueezing2,MechSqueezing3,MechEnt}. 
However, despite this success, the dissipative 
preparation of mechanical  pure states is currently limited by the linear character of the evolution, which restricts the 
set of target states to Gaussian ones~\cite{GaussRev,MatteoRev}.
\par
In order to prepare  non-Gaussian---and especially nonclassical---states of motion, some source of nonlinearity is 
needed~\cite{Note}. Early proposals for generating mechanical nonclassical states in optomechanical systems exploited 
the regime of single-photon strong coupling~\cite{Bose97,Marshall}, which however is extremely weak in current 
experimental platforms. Conditional strategies have also been  developed, e.g.~based on photon-subtraction or pulsed 
interactions, which however suffer from being probabilistic and/or having a low efficiency
~\cite{Mauro1,Mauro2,Oriol,Andersen,Vanner1,Vanner2}. In contrast, reservoir engineering guarantees the stable and 
unconditional preparation of the desired state.
\begin{figure}[t!] 
\includegraphics[width=1\columnwidth]{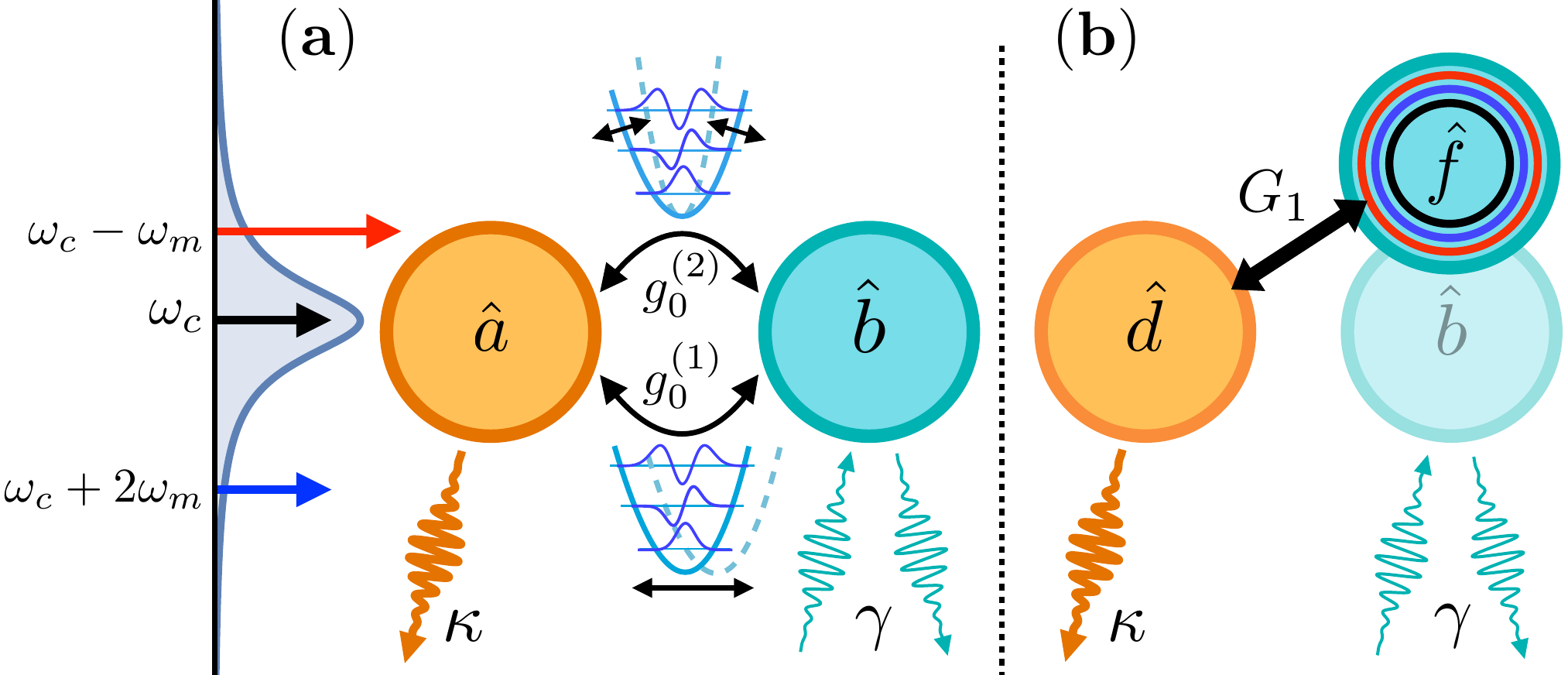}
\caption{
({\bf a}) A cavity mode ($\hat a$) and a mechanical oscillator ($\hat b$)
are coupled via a linear-and-quadratic optomechanical interaction with strength $g_0^{(1,2)}$ [Eq.~\eqref{HInt}]. 
The cavity is driven 
with three lasers as shown on the left side.  ({\bf b}) The cavity fluctuation $\hat d$ is coupled via a beam-splitter interaction 
(with strength $G_1$) to the operator $\hat f$, which is a nonlinear function of  $\hat b$ [Eqs.~\eqref{BS} and~\eqref{f}] and 
whose form is determined by the relative strengths and phases among the drives (symbolized by the circles). The prevailing cavity 
losses, which couple the system at a rate $\kappa$ to an environment with zero thermal occupation, drive the oscillator toward the 
desired state [Eqs.~\eqref{Solution} or~\eqref{NewState}] while mechanical damping at a rate $\gamma$ introduces imperfections 
[see Fig.~\ref{f:PlotFidelity}]. 
\label{f:Sketch}}
\end{figure}
\par
In this Letter we propose a dissipative scheme that exploits both the linear and the nonlinear (quadratic) optomechanical coupling 
between one cavity mode and one mechanical resonator to generate highly nonclassical states of motion of the mechanical element. 
In our scheme, the cavity provides a tunable reservoir whose properties are controlled by applying  three coherent drives.
A specific choice of their relative strengths  and phases yields a novel class of bosonic steady states that admits a simple analytical 
expression. These states are (squeezed and displaced) finite superpositions of phonon number states with fixed parity and are 
parametrized by a non-negative integer $n$, which determines how many number states are superimposed. By selecting $n=1$ we 
can stabilize  a (squeezed displaced) single-phonon state, while for increasing $n$ the state becomes a macroscopic quantum 
superposition similar to a Schr\"{o}dinger cat state. Our scheme thus interpolates between the two prototypes of nonclassicality for 
bosonic systems: from single-excitation nonclassicality, revealed in the phase space by a single pronounced negativity of the Wigner 
function, to ``interference fringes'' typical of macroscopic superposition states. These features are shown to be robust against the effect 
of mechanical dissipation. 
\par
Contrary to existing proposals for the dissipative preparation of  Schr\"{o}dinger cats  that rely on a purely quadratic optomechanical 
coupling~\cite{OptoCat1,OptoCat2}, our scheme does not require initialization to the ground state, given that the target state corresponds 
to a unique steady state.  Our proposal also differs from that of Refs.~\cite{FockNonlin,OptoCat3} inasmuch as it does not require  any 
anharmonicity of the potential. Finally, our protocol does not rely on the prohibitive single-photon strong coupling, which has 
been exploited to stabilize mechanical single phonon states~\cite{DissSingleFock} and certain 
sub-Poissonian states~\cite{DissSubP}.

\par  
\section{Model}
We consider a cavity mode whose frequency is parametrically coupled to the displacement and the displacement squared 
of a mechanical resonator. The Hamiltonian is given by (we set $\hbar=1$ throughout)
\begin{equation} \label{HInt}
\hat H=\omega_c \hat a^\dag \hat a+\omega_m \hat b^\dag \hat b - g_0^{(1)}\hat a^\dag \hat a (\hat b+\hat b^\dag)- g_0^{(2)} \hat a^\dag 
\hat a (\hat b+\hat b^\dag)^2,
\end{equation}
where $\hat a$ ($\hat b$) describes the cavity (mechanical) mode with frequency $\omega_c$ ($\omega_m$) and $g_0^{(1,2)}$ 
quantifies  the single-photon coupling strengths~\cite{OptoRev}.
We will refer to the term in Eq.~\eqref{HInt} proportional to the mechanical position (position squared) as the linear (quadratic) term; as 
sketched in Fig.~\ref{f:Sketch} ({\bf a}), its action consists in the displacement (squeezing) of the mechanical mode conditioned on the
number of cavity photons.  
\par
The cavity is driven with three lasers, one red-detuned by one mechanical frequency, one blue-detuned by twice the mechanical frequency 
and one resonant, as schematically shown in Fig.~\ref{f:Sketch} ({\bf a}).  
The effect of the drives is taken into 
account by the displacement transformation  $\nobreak{\hat a=\sum_k \alpha_k e^{-i \omega_k t}+\hat d}$, where $\alpha_k$ is the intra-cavity 
amplitude at each driving frequency $\omega_k$ and $\hat d$ is a quantum fluctuation.  Moving to a frame rotating with the cavity and mechanical 
frequencies, we can write the displaced Hamiltonian as $\hat H=\hat H_{{\rm RWA}}+\hat H_{{\rm CR}}$, where $\hat H_{{\rm RWA}}$ contains 
the transitions resonantly enhanced by the drives while $\hat H_{{\rm CR}}$  collects the off-resonant terms.
If we restrict ourselves to the limit 
$|G_{1,2,3}|\ll\omega_m,\, |R G_1|\ll\omega_m$  and $|R^{-1} G_{2,3}|\ll\omega_m$, where $R=g_0^{(2)}/g_0^{(1)}$, 
$G_1=\alpha_1g_0^{(1)}$ and $G_{2(3)}=\alpha_{2(3)}g_0^{(2)}$,
we can neglect the counter-rotating terms and consider only the resonant contributions (cf. Appendix~\ref{app:Derivation})
\begin{equation}\label{BS}
\hat H_{{\rm RWA}}=G_1 (\hat d^{\dag}\hat f +\hat d\, \hat f^{\dag})  \, ,
\end{equation}
where we have introduced the  operator
\begin{equation}\label{f}
\hat f= \hat b+\frac{G_2}{G_1} \hat b^{\dag\,2}+\frac{G_3}{G_1}\bigl(\hat b \hat b^{\dag}+\hat b^{\dag} \hat b\bigr) \, .
\end{equation}
In the following we will take the coefficients $G_{1,2,3}$ to to be real without loss of generality.  Eq.~\eqref{BS} describes a beam-splitter 
interaction between the cavity fluctuation and a nonlinear combination of the mechanical creation and annihilation operators, as shown 
in Fig.~\ref{f:Sketch} ({\bf b}).  
The form of  Eq.~\eqref{f} stems from the joint presence 
of the linear and the quadratic coupling between one cavity mode and {\it one} mechanical oscillator;
coupling to {\it different} cavity modes have been recently considered to obtain a tunable optomechanical nonlinearity~\cite{TunableNonLin}.
\par
We also need to take into account the effect of dissipation. 
We start by including the dominant cavity losses, in which case  the evolution of the joint density matrix $\hat\varrho$ reads
\begin{equation}\label{MasterEq}
\dot{\hat \varrho}=-i[\hat H_{{\rm RWA}},\hat \varrho]+\kappa\, \mathcal{D}_{d}[\hat \varrho] \, ,
\end{equation}
where $\mathcal{D}_{ o}[\hat\varrho]=\hat o \hat \varrho \hat o^{\dag}-\frac12\bigl(\hat o^{\dag} \hat o\hat \varrho + 
\hat \varrho\hat o^{\dag} \hat o \bigr)$ is the standard dissipator. Provided that a stationary state exists, this is given by 
$\hat \varrho_{ss}=\ket{\psi_{ss}}\bra{\psi_{ss}}$, with $\ket{\psi_{ss}}=\ket{0}\otimes\ket{\varphi}$ and where the mechanical state obeys 
the dark state condition~\cite{Kraus} 
\begin{equation}\label{DarkState}
\hat f \ket{\varphi}=0 \, .
\end{equation}
By varying the number, strength and frequency of the drives, reservoir engineering with 
linear-and-quadratic coupling allows to stabilize a plethora of nonclassical states and manifolds thereof~\cite{Us}. In the 
following we focus on a particularly relevant instance. 
\par
\emph{Novel family of steady states.}---We now introduce and characterize a new family of states that are generated within 
the scheme presented above. If we assume  
\begin{equation}\label{Choice}
G_3=- G_2=\frac{G_1}{2\sqrt{2n+1}} \, ,
\end{equation} 
where  $n\in\mathbb{N}_0$ is a non-negative integer, the mechanical steady state $\ket{\varphi}\equiv\ket{\varphi_n}$ is described 
by the surprisingly simple wave function 
\begin{equation}\label{Solution}
\varphi_n(x)\propto e^{-\frac{X_n^2}{4}} H_n(X_n)\, .
\end{equation}
In the equation above, $\varphi_n(x)=\langle x\ket{\varphi_n}$ and  $H_n(X_n)$ is the Hermite polynomial of  argument 
$\nobreak{X_n=\sqrt{\frac{2}{3}}\left(x+\sqrt{4n+2}\right)}$.
\begin{figure*}[t]
	\centering
\includegraphics[width=\linewidth]{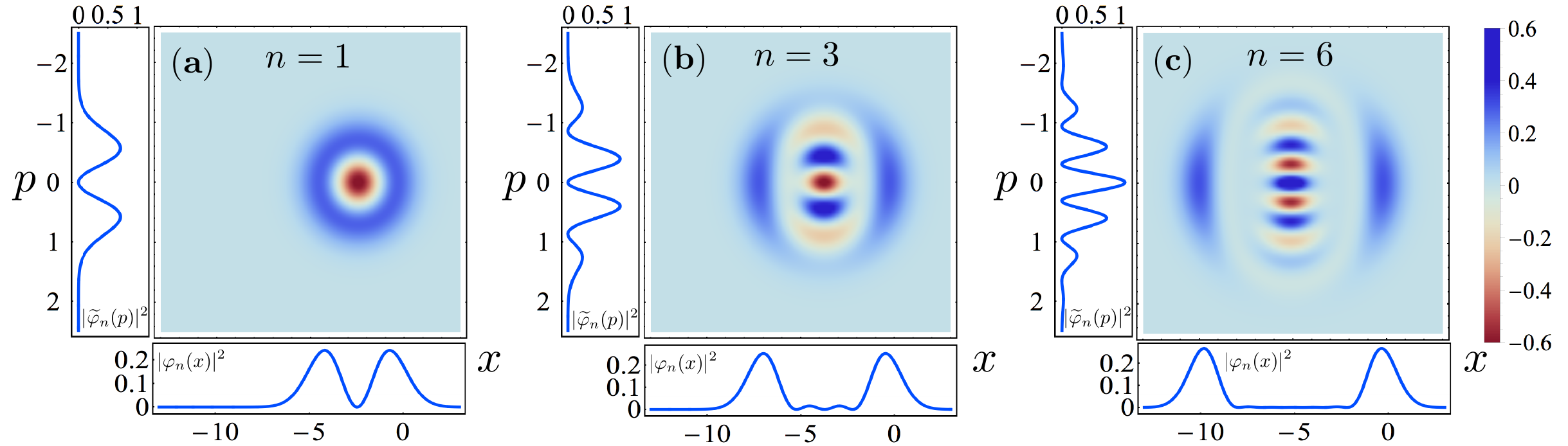}
\caption{Wigner function $W(x,p)$ of the state $\ket{\varphi_n}$ [Eq.~\eqref{NewState}] for $n=1$ ({\bf a}),  $n=3$ ({\bf b})
and  $n=6$ ({\bf c}). The marginals of $W(x,p)$, which provide the position and momentum probability distribution, are also shown. 
\label{f:PlotWigner}}
\end{figure*}
This expression has been obtained 
by solving the  differential equation associated 
Eq.~\eqref{DarkState}~(cf. Appendix~\ref{app:Solution}).
 The choice of the coupling strengths  as in Eq.~\eqref{Choice}, and in particular the 
introduction of an integer parameter, are crucial to obtain such a simple expression. Nevertheless, we verified numerically  that 
for small deviations from these  values, the steady state (now no longer pure) has near-unit fidelity with the target state described 
by Eq.~\eqref{Solution}, so that no fine-tuning issue arises.
\par
The stationary wave function $\varphi_n(x)$  resembles that of a simple harmonic oscillator, however with  two crucial differences: 
(i) the integer $n$ appears 
both in the order and in the argument of the Hermite polynomial
and (ii) the presence 
of a factor 4 in the exponential. The  latter, albeit seemingly innocuous,  prevents $\varphi_n(x)$ from being recast into the standard 
harmonic oscillator form  and, in fact,  entails 
a superposition of harmonic oscillator wave functions.
The  corresponding state is indeed 
a squeezed and 
displaced  superposition of a finite number of Fock states~(cf. Appendix~\ref{app:Solution}) 
\begin{equation}\label{NewState}
\ket{\varphi_n}\propto \hat D(\zeta_n) \hat S(r)\sum_{j=0}^{\lfloor \frac{n}{2}\rfloor}\frac{1}{2^{2j}j!\sqrt{(n-2j)!}}\ket{n-2j} \, ,
\end{equation}
where $\hat S(r)$ and $\hat D(\zeta_n)$ are the squeezing and displacement operator of argument $r= \ln \sqrt3$ and 
$\nobreak{\zeta_n~=-\sqrt{2n+1}}$, respectively, and $\lfloor y \rfloor$ yields the greatest integer smaller or equal than $y$. 
Eq.~\eqref{NewState} provides the exact expression of a new instance of a bosonic state and represents one of the main 
results of this work. Unlike 
coherent or squeezed states, for which the 
coefficients are found by writing the definition [analogue of Eq.~\eqref{DarkState}] in the Fock basis and solving a recurrence 
relation, such an attempt here would fail. Instead, our approach of first obtaining the wave function by projecting 
the dark state condition onto the position eigenstates and from that deriving a closed expression for the coefficients proves 
successful. 
\par
The state $\ket{\varphi_n}$ consists of two Gaussian unitary operations acting on a {\it finite} superposition of Fock states,
which is responsible for its nonclassical nature. This finite seed contains at most $n$ excitations, has a definite number 
parity and  can be in principle isolated by deterministically counter-squeezing and displacing the state. 
Theoretical proposals to 
achieve probabilistically the truncation of photon number superpositions have been put forward for linear optical devices
~\cite{QuantumScissors1,QuantumScissors2}. In contrast, here a finite superposition can be obtained unconditionally, without 
exploiting entanglement and for a massive system. These states may thus be  useful  for quantum information processing as a 
robust choice for qubit encoding~\cite{Gottesman}, similarly to what has already been proposed for Schr\"{o}dinger 
cat states~\cite{Cat1,Cat2}.     
\par 
In Fig.~\ref{f:PlotWigner} we show the Wigner function 
$\nobreak{W(x,p)=\frac{1}{\pi}\int\mathrm{d}y\, e^{-2ipy}\varphi_n(x+y)\varphi_n^*(x-y)}$ 
of $\ket{\varphi_n}$
for different values of $n$. The transition from a single pronounced negativity 
({\bf a}) to phase-space ``ripples'' ({\bf b})-({\bf c}) is apparent. It is useful to compare our solution to the family of Schr\"{o}dinger 
cat states $\ket{\mathcal{C}_{\alpha}^\pm}\propto\ket{\alpha}\pm\ket{-\alpha}$~\cite{WallsMilburn}, 
for which optomechanical realizations exploiting  reservoir engineering have been proposed
~\cite{OptoCat1,OptoCat2,OptoCat3}. Contrary to the case of an odd cat state $\ket{\mathcal{C}_{\alpha}^-}$, which in the limit 
of small amplitude {\it approximates} a single-phonon state---the so-called ``kitten'' state~\cite{Kitten}---the state 
$\ket{\varphi_1}=\hat D(\zeta_1)\hat S(r)\ket{1}$ is \emph{exactly} a (squeezed and displaced) single-phonon state. On the other 
hand, for large $n$ the state $\ket{\varphi_n}$ approaches a Schr\"{o}dinger cat, yet the two never fully overlap (even asymptotically 
unit fidelity is not attained), so that Eq.~\eqref{NewState} embodies a similar but distinct instance of a macroscopic quantum 
superposition~(cf. Appendix~\ref{app:Comparison}).
\par  
\section{Rate of approaching the steady state and effects of mechanical dissipation}
We now address how the unavoidable 
presence of mechanical damping affects the properties of the target state. For simplicity, we focus on the fast cavity limit 
$\kappa\gg G_k$, where adiabatic elimination of the cavity field leads to an effective master equation for the  reduced 
mechanical density matrix (see Refs.~\cite{QuantumNoise,Wiseman} or cf. Appendix~\ref{app:effective-master-equation}
for explicit derivation)
\begin{equation}\label{Adiabatic}
\dot{\hat \varrho}^{(m)}=\gamma \mathcal{C}\,\mathcal{D}_{ f}\,\bigl[\hat \varrho^{(m)}\bigr]+
\gamma (\bar n+1)\mathcal{D}_{ b}\bigl[\hat \varrho^{(m)}\bigr]+\gamma \bar n \mathcal{D}_{ b^\dag}\bigl[\hat \varrho^{(m)}\bigr] \,,
\end{equation}
where $\mathcal{C}=4G_1^2/(\gamma\kappa)$ defines the optomechanical cooperativity. The first term on the right-hand 
side describes dissipation induced by the modified jump operator
\begin{equation}\label{Modef}
\hat f=\hat b-\frac{1}{2\sqrt{2n+1}}\bigl[\hat b^{\dag\,2}-\bigl(\hat b \hat b^{\dag}+\hat b^{\dag} \hat b\bigr)\bigr] \, ,
\end{equation}
which makes manifest the role played by the cavity in providing an engineered environment for the mechanical degree of 
freedom. In Eq.~\eqref{Adiabatic} we also added  thermal decoherence  to a mechanical bath at a rate $\gamma$ and 
with $\bar n$ thermal occupancy.
\par
Let us first consider the limit of no mechanical damping. 
In this case Eq.~\eqref{Adiabatic} describes a purely dissipative 
dynamics, however relative to a jump operator that is {\it neither linear nor bosonic}; 
complete information about the dynamics can be uncovered by studying the spectrum of $\mathcal{D}_{ f}$.
In the infinite-time limit the state $\hat \varrho_{ss}^{(m)}=\lim_{t\rightarrow\infty}\hat\varrho^{(m)}$ 
satisfies $\mathcal{D}_{f}\bigl[\hat \varrho_{ss}^{(m)}\bigr]=0$ and is non-degenerate. 
We can conclude that our protocol is both deterministic and independent of the choice of the initial state, allowing in 
principle to start from any given state, e.g.~a thermal one. This must be contrasted with dissipative preparation of 
mechanical cat states~\cite{OptoCat1,OptoCat2}, for which the steady state has a double degeneracy and consequently 
initialization to a state of definite parity---typically the ground state---is needed. The presence of a 
linear part in $\hat f$ breaks the discrete parity symmetry associated with the quadratic terms and makes the system more robust 
to losses~\cite{Symmetries}.  We also stress that our method enables the preparation of macroscopic superposition 
states of chosen parity, while reservoir engineering of odd cat states is highly impractical, as it requires to initialize the 
state to a pure odd-parity state, e.g.~in $\ket{1}$.
\begin{figure}[t!] 
\includegraphics[width=1.0\columnwidth]{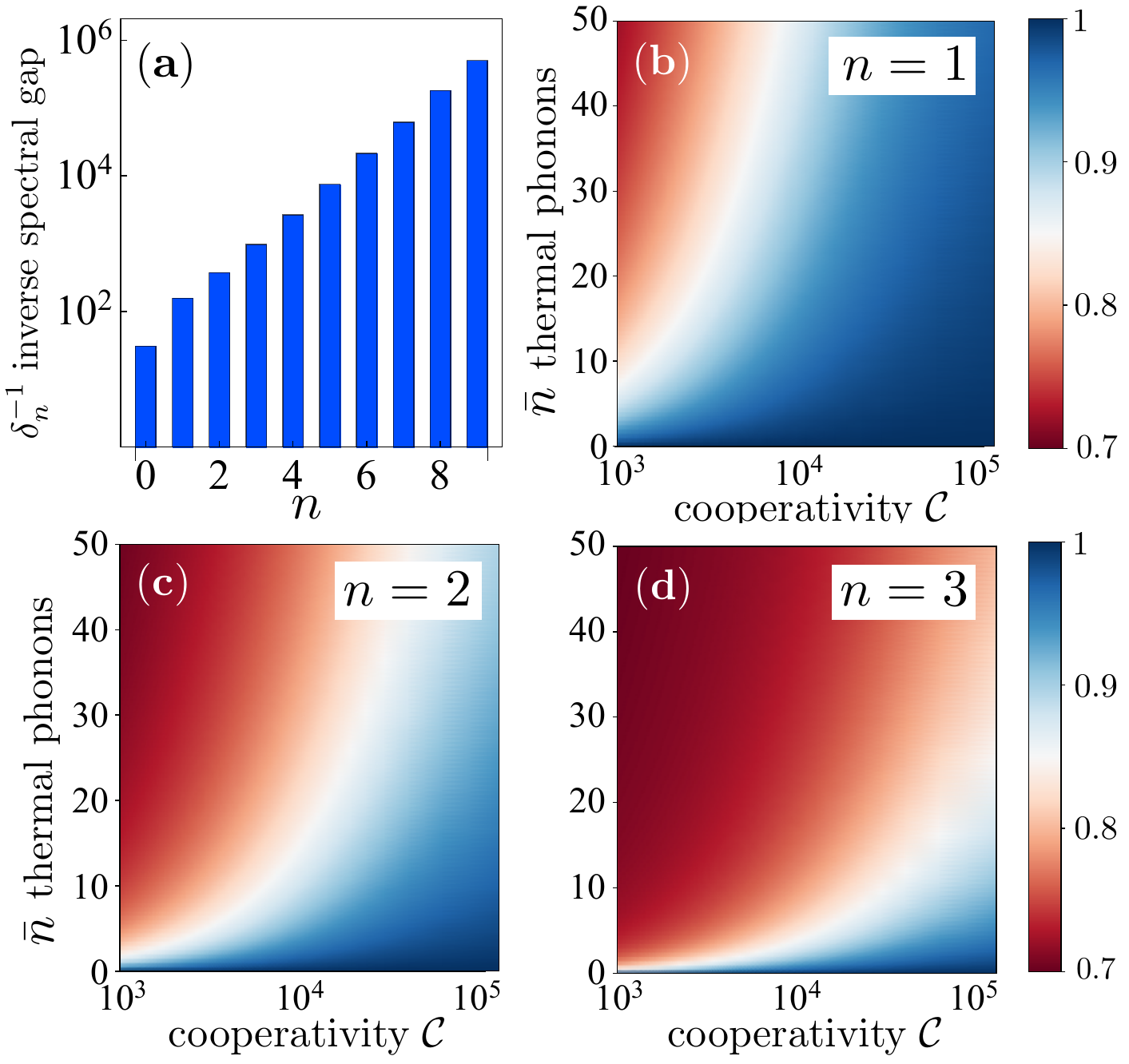}
\caption{({\bf a}) Plot of the inverse spectral gap $\delta_n^{-1}$ of the dissipator $\mathcal{D}_{ f}$ with jump operator 
as in Eq.~\eqref{Modef} for different $n$, in units of $(\gamma\mathcal{C})^{-1}$. $\delta_n/(\gamma\mathcal{C})=-\mathrm{Re}\, 
\lambda_n$, with $\lambda_n$ the smallest non-zero eigenvalue. 
({\bf b})-({\bf d}) Fidelity between the target state Eq.~\eqref{NewState} and the steady state in the presence 
of mechanical damping as a function of $\bar n$ and $\gamma$ (parametrized by the cooperativity $\mathcal{C}$) for 
$n=1$ ({\bf b}),  $n=2$ ({\bf c}) and  $n=3$ ({\bf d}). We set $\nobreak{G_1=0.05\kappa}$.
\label{f:PlotFidelity}}
\end{figure}
\par
The analysis of the spectrum also provides information about the timescale required to reach the steady state. 
The slowest decaying term is associated with the eigenvalue $\lambda_n$ that has the smallest non-zero real part, which 
in turn defines the spectral gap $\delta_n/(\gamma\mathcal{C})=-\mathrm{Re}\, \lambda_n$. The time needed to approach the steady state
(with a fixed fidelity) scales as $\nobreak{\tau_n\sim \delta_n^{-1}}$. As pointed out, when $n$ increases the state $\ket{\varphi_n}$ 
realizes a macroscopic superposition. On the other hand, from Eq.~\eqref{Modef} we see that for increasing $n$ the 
nonlinear terms are suppressed with respect to the cooling part: the optimal strategy to prepare a large quantum 
superposition is therefore to minimally perturb---in the way prescribed by Eq.~\eqref{Modef}---a standard cooling process, 
which is recovered by setting~$G_2=G_3=0$ in Eq.~\eqref{f}. However, we find that the spectral gap is exponentially 
suppressed with respect to $n$  and therefore  $\tau_n$ grows exponentially, as shown in Fig.~\ref{f:PlotFidelity} ({\bf a}). 
It also 
represents the main limiting factor of our protocol for the generation of macroscopic quantum states when thermal 
decoherence is taken into account. 
\par
We finally introduce a nonzero coupling with the bath.  Figs.~\ref{f:PlotFidelity} ({\bf b})-({\bf d}) show the
mismatch, quantified by the fidelity~\cite{Uhlmann}, between the actual steady state and the target state in 
Eq.~\eqref{NewState}. As expected, mechanical dissipation is responsible for a decrease of the purity and states with 
greater $n$ are more susceptible to thermal decoherence. Nevertheless, we see that regions of near-unit fidelity are 
present even for considerable thermal occupancy. Moreover, even if the fidelity is no longer close to one, we show 
that the steady state retains nonclassical features and is always non-Gaussian (in the range of parameters 
explored, see Appendix~\ref{app:NonClassVolume}).  
\par
\section{Experimental implementation}
To implement our idea,
we consider a three-mode optomechanical system made of a pair of 2D photonic crystal cavities in a double-slotted configuration separated
by a central mechanical beam~\cite{NonLin5,NonLin6}. When the beam is equidistant from the two slabs, an enhanced quadratic 
optomechanical coupling is obtained~\cite{QND2}. On the other hand, a shift in the beam's position, which can be controlled via electrostatic 
actuation, determines a tunable coupling that has both a linear and a quadratic component~(cf. Appendix~\ref{app:Implementation}).  
To estimate the single-photon couplings, we consider the following realistic parameters from Ref.~\cite{NonLin6}: photon tunneling rate between left and right cavity 
$J/2\pi=0.1$ GHz, bare frequency pull parameter  $g_L=-g_R=100$ GHz/nm, zero point amplitude of the nano-beam $x_\mathrm{zpf}=10$ fm, 
application of a bias voltage of a few tenth of millivolts (that guarantees a displacement $x_0<0.1$ pm). 
These parameters yield $g_0^{(2)}\approx 5$ kHz and $g_0^{(1)}\approx 70$ kHz,
and hence a ratio $R\approx0.07$, for which the RWA in Eq.~\eqref{BS} is justified~(see Appendix~\ref{app:Derivation} for details).
The central beam supports  several acoustic modes
ranging from a few MHz to a few GHz, with
modes of frequency $\omega_m > 300$ MHz lying deep in the resolved-sideband regime. 
Photonic crystal cavities allow for large intra-cavity photon capacities $n_c>10^4$, which give couplings 
$G_{1,2,3}$  in the 10 -- 100 MHz range. The reasonable choice of the parameters $Q_m=10^6$, $\omega_m/2\pi=400$ MHz, $\kappa/2\pi=50$ MHz, 
$n_c=10^4$, $g_0^{(1)}=70$ kHZ gives a multiphoton cooperativity $\mathcal{C}\approx 10^4$, which can be easily increased by one order of magnitude
by considering higher mechanical $Q$ and/or higher couplings.
This would allow for the stabilization of our target state with high fidelity also without initialization in the ground state.
\par
Once the target state $\hat \varrho_{ss}^{(m)}$ has been prepared, a cavity mode  (different from the one providing the engineered reservoir)
can be employed for the readout. Tomographic schemes via quantum non-demolition (QND) coupling have been proposed both in the good~\cite{BAE,Darren} 
and bad cavity limit~\cite{Pulsed1,Pulsed2,NoCavity,TomographyRev}, and can be 
directly apply here. A less demanding task would be the certification of the nonclassicality of the state, which can be accomplished  with a single 
homodyne-like measurement~\cite{NoCavity}. In photonic crystal architecture, it may be especially convenient to exploit the optomechanical interaction with one 
of the two cavity supermodes for the preparation and the other for the readout.  
\par
\section{Conclusions}
The linear and the quadratic couplings achievable  in optomechanical systems have so far been 
addressed separately. We showed that the joint presence of both terms enables  engineering of unique nonclassical 
features in the state of a mechanical resonator.
Our proposal 
achieves the unconditional preparation of states of a macroscopic object featuring a non-positive Wigner function. 
\par
\section{Acknowledgements}
We thank M. Sillanp\"a\"a and T. K. Para\"iso for helpful discussions.
This work was supported by the European Union's Horizon 2020 research and innovation programme under grant 
agreement No 732894 (FET Proactive HOT). A.~N.~acknowledges a University Research Fellowship from the Royal 
Society and additional support from the Winton Programme for the Physics of Sustainability. D.~M.~acknowledges 
the Coordinator Support funds from Queen's University Belfast. O.~H.~and M.~P.~acknowledge support from the 
SFI-DfE Investigator programme (grant 15/IA/2864). O.~H., A.~F., and M.~P.~are supported by the EU Horizon2020 
Collaborative Project TEQ (grant agreement nr. 766900). A.~F.~and O.~H.~acknowledge funding from the EPSRC 
project EP/P00282X/1. M.~P.~thanks COST Action CA15220 QTSpace for partial support.





\begin{appendix}
\section{Derivation of the Hamiltonian 
and the effect of counter-rotating terms}\label{app:Derivation}
We consider an optomechanical system where the frequency of a cavity mode parametrically couples to the 
displacement and  the displacement squared of a mechanical resonator~\cite{OptoRev}. The Hamiltonian is 
given by $(\hbar=1)$
\begin{equation}
\hat H=\hat H_0+\hat H_{\text{int}}+\hat H_{\text{drive}} \label{SHTot} \, ,
\end{equation}
where we set
\begin{subequations}
\begin{align}
\hat H_0&=\omega_c \hat a^\dag \hat a+\omega_m \hat b^\dag \hat b \label{SH0}\, , \\ 
\hat H_{\text{int}}&=- g_0^{(1)} \hat a^\dag \hat a (\hat b+\hat b^\dag)- g_0^{(2)} \hat a^\dag \hat a 
(\hat b+\hat b^\dag)^2 \label{SHInt}\, , \\
\hat H_{\text{drive}}&=\mathcal{E}(t) \hat a^\dag+\mathcal{E}^*(t) \hat a \label{SHDrive} \, .
\end{align}
\end{subequations}
The first of these expressions, Eq.~\eqref{SH0}, contains the free oscillating terms, where 
$\hat a$ ($\hat b$) describes the cavity (mechanical) mode with frequency $\omega_c$ ($\omega_m$).
The second, Eq.~\eqref{SHInt}, describes the linear and the quadratic optomechanical interaction with single-photon 
coupling strength $\nobreak{g_0^{(k)}=-x_{\text{zpf}}^k 2^{1-k}\partial^k_x\omega_c(\hat x)\vert_{x=0}}$, 
$k=1,2$, $\hat x$ being the (dimensionless) mechanical displacement and $x_{\text{zpf}}$ the zero-point 
fluctuation. The last expression, Eq.~\eqref{SHDrive}, includes a coherent drive of the cavity with multiple 
tones of frequency $\omega_k$ and amplitude $\epsilon_k$, namely 
$\mathcal{E}(t)=\sum_k \epsilon_k e^{-i \omega_k t}$. 
\par
The cavity is in contact with an effective zero-temperature reservoir provided by the extra-cavity modes, 
while the mechanical oscillator is in contact with a bath of inverse temperature $\beta$ that induces 
$\bar n=(e^{\beta \omega_m}-1)^{-1}$ average thermal excitations~\cite{WallsMilburn,QuantumNoise}. 
We will assume for both processes the Markovian limit, that translates into the following expressions for the 
correlation functions  of the optical ($\hat{a}_{\text{in}}$) and mechanical ($\hat{b}_{\text{in}}$) input noise operator   
\begin{subequations}
\begin{align}\label{SNoiseOpt}
\langle \hat{a}_{\text{in}}(t)\hat{a}_{\text{in}}^{\dagger}(t')\rangle&= \delta (t-t')\, ,
&\langle \hat{a}_{\text{in}}^{\dagger}(t)\hat{a}_{\text{in}}(t')\rangle&=0 \, ,
\\
\langle \hat{b}_{\text{in}}(t)\hat{b}_{\text{in}}^{\dagger}(t')\rangle&= (\bar n +1) \delta (t-t')\, ,
&\langle \hat{b}_{\text{in}}^{\dagger}(t)\hat{b}_{\text{in}}(t')\rangle&=\bar n \,\delta (t-t')\, .
\end{align}   
\end{subequations}
The Heisenberg-Langevin equations for the system are thus given by
\begin{subequations}
\begin{align}
\dot{\hat a} &=-i\left[ \omega_c-g_0^{(1)}(\hat b+\hat b^\dag)-g_0^{(2)}(\hat b+\hat b^\dag)^2 \right] \hat a 
-\frac{\kappa}{2}\hat a-i\mathcal{E} \nonumber \\
&+\sqrt{\kappa} \hat a_{\mathrm{in}} \, , \label{SOptLangevin} \\
\dot{\hat b} &=-i\omega_m \hat b+ i\left[g_0^{(1)}+2 i g_0^{(2)}(\hat b+\hat b^\dag)\right]\hat a^\dag a 
-\frac{\gamma}{2}\hat b+\sqrt{\gamma} \hat b_{\mathrm{in}} \, , \label{SMechLangevin}
\end{align}
\end{subequations}
where $\kappa$ and $\gamma$ are the optical and the mechanical damping rate.
\par
We then separate the contributions to the dynamics into mean field and  fluctuations, 
i.e.~$\nobreak{\hat a(t)=\alpha(t)+\hat d(t)}$. After a transient, we expect the cavity field to follow the 
modulation of the drive, i.e.~$\alpha(t)=\sum_k \alpha_k e^{-i \omega_k t}$. Driving multiple frequencies 
leads to amplitude modulation of the intra-cavity field, which in turn translates into an oscillating force acting 
on the mechanical element. This fact can be taken into account by decomposing also the mechanical mode 
into mean field and  fluctuations, $\hat b(t)=\beta(t)+\hat h(t)$. Furthermore, if we restrict ourselves to 
the limit $g_0^{(j)} \alpha_k \alpha_l\ll \omega_m$, $j=1,2$, the mean fields attain a stationary value, which 
we refer to as $\alpha_{k,s}\,, \beta_s$. The steady amplitudes take  the following expressions  
\begin{subequations}
\begin{align}
\alpha_{k,s} &=\frac{-i \epsilon_k}{\frac{\kappa}{2} -i \left[\Delta_k+g_0^{(1)}(\beta_s+\beta_s^*)
+g_0^{(2)}(\beta_s+\beta_s^*)^2 \right] } \, , \\
\beta_s &=\frac{g_0^{(1)}  \sum_k \left\vert\alpha_{k,s} 
\right\vert^2\left( \omega_m+i\frac{\gamma}{2}\right)}{\left(\frac{\gamma}{2}\right)^2
+\omega_m\left(\omega_m-4g_0^{(2)}\sum_k \left\vert\alpha_{k,s} \right\vert^2 \right)} \, , \label{5}
\end{align}
\end{subequations}
where we set $\Delta_k=\omega_k-\omega_c$. We see that position and position-squared couplings lead to 
a shift of the equilibrium mechanical position and to a modified detuning. However, these effects are small 
and can be safely neglected, therefore we set 
$\beta_s\approx 0$ and $\alpha_{k,s} =\frac{-i \epsilon_k}{\kappa/2 -i \Delta_k }$ in what follows.
\par
Moving to an interaction picture with respect  to $\hat H_0$ the Hamiltonian is transformed into
\begin{widetext} 
\begin{equation}\label{SHLin}
\hat H=-\sum_k \left(\alpha_k  \hat d^{\dag}e^{-i\Delta_k t}+\alpha_k^* \hat d e^{i\Delta_k t}\right)
\left[g_0^{(1)}\left(\hat b e^{-i\omega_m t}+\hat b^{\dag} e^{i\omega_m t}\right)+g_0^{(2)}
\left(\hat b e^{-i\omega_m t}+\hat b^{\dag} e^{i\omega_m t}\right)^2 \right] \, .
\end{equation}
\end{widetext}
We now consider the following choice for the drives
\begin{align}\label{SDrives}
\Delta_1&=-\omega_m\, , \hspace{-1cm} &\Delta_2&=2\omega_m\, ,\hspace{-1cm} &\Delta_3&=0\, ,
\end{align}
which correspond to driving the first red mechanical sideband, the second blue sideband and on the 
cavity resonance. This choice is to be understood {\it a posteriori}, as a suitable modification of a 
cavity cooling scheme that selects the nonlinear terms necessary to prepare the desired state. 
Indeed, for particular values of the strength and phase of the second and third drive with respect to 
the cooling beam, this setup cools the mechanical mode toward a nonclassical state of motion. 
The application of the drives displayed in Eq.~\eqref{SDrives} makes the following processes in the  
Hamiltonian Eq.~\eqref{SHLin} resonant  
\begin{equation}\label{SRWA}
\hat H_{{\rm RWA}}=G_1 (\hat d^{\dag}\hat f +\hat d\, \hat f^{\dag})  \, ,
\end{equation}
where 
\begin{equation}\label{SFMode}
\hat f= \hat b+\frac{G_2}{G_1} \hat b^{\dag\,2}+\frac{G_3}{G_1}\{\hat b,\hat b^{\dag}\}  \, ,
\end{equation}
and we set $G_1=\alpha_1g_0^{(1)}$, $G_{2(3)}=\alpha_{2(3)}g_0^{(2)}$ and $\{\cdot,\cdot\}$ is the anticommutator.
\begin{figure}[t!] 
\includegraphics[width=1.\columnwidth]{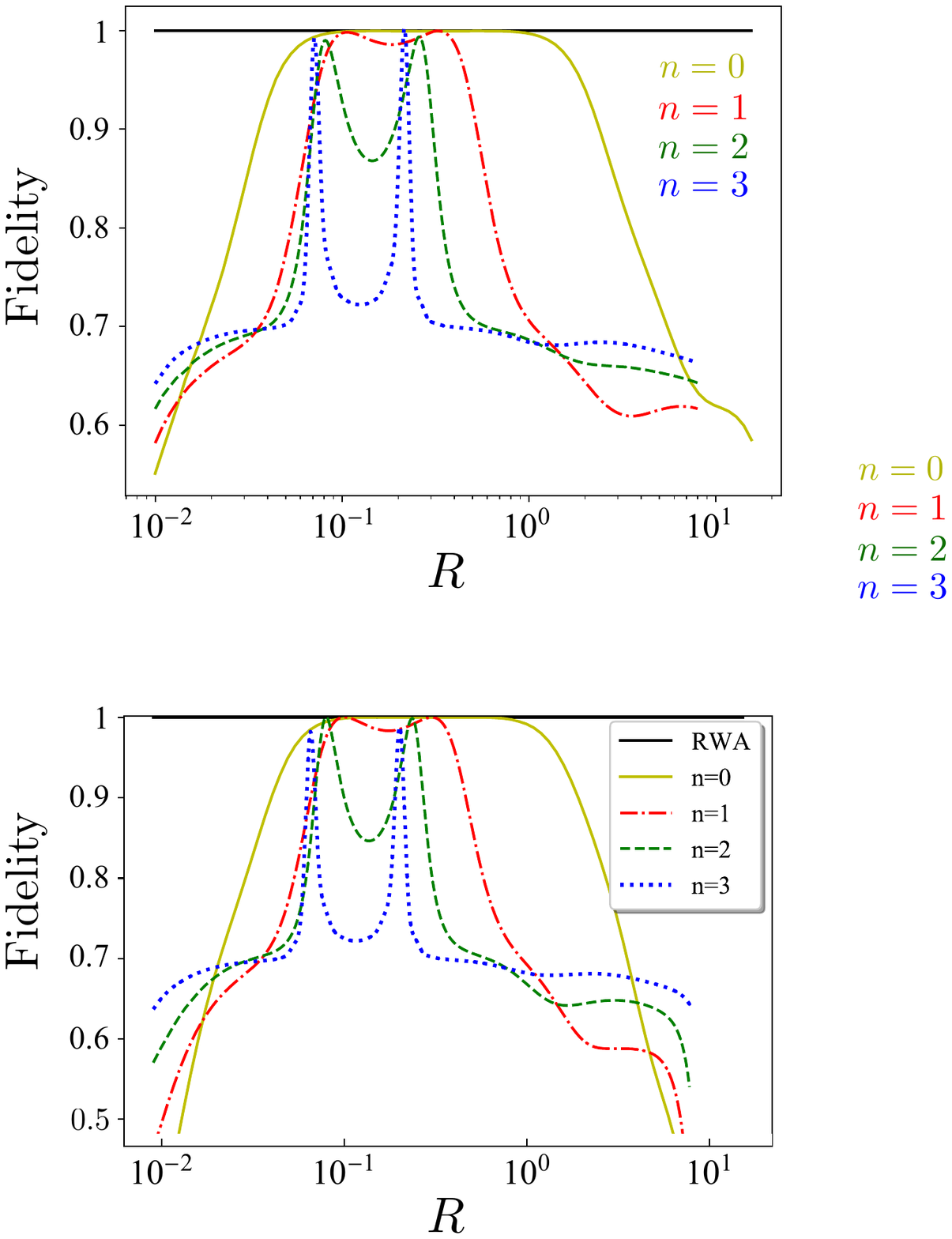}
\caption{Fidelity between the long-time average of the full master equation \eqref{FullMasterEq}
and  the target state \eqref{NewState} as a function of  the ratio $R=g_0^{(2)}/g_0^{(1)}$ between 
single-photon quadratic and linear coupling. Parameters are $\kappa = 0.001\omega_m$, 
$G_1= 0.05\kappa$. The horizontal black line corresponds to take the rotating-wave approximation, the yellow 
solid, red dot-dashed, green dashed and blue dotted curves are for $n=0,1,2,3$, respectively.
\label{f:PlotCR}}
\end{figure}
The resonant contributions Eq.~\eqref{SRWA} describe a beam-splitter interaction between the fluctuation of 
the cavity field $\hat d$ and the nonlinear combination of mechanical creation and annihilation operators $\hat f$. 
\par
The counter-rotating terms $\hat H_{{\rm CR}}=\hat H-\hat H_{{\rm RWA}}$ are 
\begin{align}\label{SCR}
\hat H_{{\rm CR}}&=\hat d^{\dag}\left\{e^{-i\omega_m t}\left(\alpha_2g_0^{(1)}  \hat b^{\dag}+\alpha_3g_0^{(1)} 
\hat b+\alpha_1g_0^{(2)} \hat b^2 \right) \right. \nonumber \\
&\left. + e^{+i\omega_m t}\left(\alpha_3g_0^{(1)}  
\hat b^{\dag}+ \alpha_1g_0^{(2)} \{\hat b,\hat b^{\dag}\} \right)\right. \nonumber \\
&\left.+ e^{-2i\omega_m t}\left(\alpha_2g_0^{(2)}  \{\hat b,\hat b^{\dag}\} + \alpha_3g_0^{(2)} 
\hat b^2 \right) \right. \nonumber \\
&\left. + e^{+2i\omega_m t}\left(\alpha_1g_0^{(1)}  \hat b^\dag 
+ \alpha_3g_0^{(2)} \hat b^{\dag\, 2} \right) \right. \nonumber \\
&\left.+ e^{-3i\omega_m t} \alpha_2g_0^{(1)} \hat b + e^{+3i\omega_m t} \alpha_1g_0^{(2)} 
\hat b^{\dag\,2} \right. \nonumber \\
&\left.+ e^{-4i\omega_m t} \alpha_2g_0^{(2)} \hat b^{ 2}\right\} 
\,+\mathrm{H.c.} \, .
\end{align}
We rewrite the oscillating terms as  $\hat H_{{\rm CR}}=\sum_{k=1}^4 e^{i\omega_m k t} \hat H_{{\rm CR}}^{(k)}+\mathrm{H.c.}$, with 
\begin{subequations}
\begin{align}
\hat H_{{\rm CR}}^{(1)}&=R^{-1} G_2 \hat d\hat b+R^{-1}\left(G_3 \hat d^\dag+G_3\hat d\right)\hat b^\dag+RG_1\hat d \hat b^{\dag\,2} \nonumber \\
&+R G_1\hat d^\dag\{\hat b,\hat b^\dag\} \, ,\\
\hat H_{{\rm CR}}^{(2)}&=G_1\hat d^\dag\hat b^\dag+\left(G_3\hat d^\dag+G_3\hat d\right)\hat b^{\dag\,2}+G_2\hat d\{\hat b,\hat b^\dag\} \, , \\
\hat H_{{\rm CR}}^{(3)}&=R^{-1} G_2\hat d\hat b^\dag+R G_1 \hat d^\dag \hat b^{\dag\,2} \, , \\
\hat H_{{\rm CR}}^{(4)}&=G_2\hat d\hat b^{\dag\, 2}\,,
\end{align}
\end{subequations}
where we introduced the ratio   $R=g_0^{(2)}/g_0^{(1)}$ between the quadratic and the linear single-photon coupling
strength. From this explicit form it is apparent that a necessary condition for the RWA to be valid is that
\begin{equation}\label{RWACondition}
\left\vert  G_{1,2,3} \right\vert \ll \omega_m \,,\quad \left|R G_1\right|\ll\omega_m \quad \mathrm{and}\quad \left|R^{-1} G_{2,3}\right|\ll\omega_m \,.
\end{equation}
We can verify the validity of the RWA by integrating numerically the time-dependent master equation 
\begin{equation}\label{FullMasterEq}
\dot{\hat \varrho}=-i[\hat H_{{\rm RWA}}+\hat H_{{\rm CR}},\hat \varrho]+\kappa\, \mathcal{D}_{d}[\hat \varrho] 
\end{equation} 
and comparing its long-time average with the steady state of the same master equation when omitting the counter rotating terms $\hat H_{{\rm CR}}$. In the following we choose the driving amplitudes such that
\begin{equation}\label{SChoice}
	G_3=- G_2=\frac{G_1}{2\sqrt{2n+1}} \, ,
\end{equation} 
where $n$ is a non-negative integer. 
In Fig.~\ref{f:PlotCR} we show the fidelity between the two steady states of the master Eq.~ \eqref{FullMasterEq}, with and without $\hat H_{\mathrm{CR}}$, as a function of
the ratio $R$ for different values of $n = 0,1,2,3$ and the parameters $\kappa = 0.001\omega_m$, $G_1= 0.01\kappa$. 
We can see that there exist ranges of values of $R$, within the region identified by the conditions in Eq.~\eqref{RWACondition}, for which the rotating-wave approximation is fully justified (for the parameters under consideration).
The range of values $R$ for which the counter-rotating terms can be neglected 
depends on $n$, and the fidelity develops a double-peak structure that shrinks for increasing $n$;
this behavior can be understood from the fact that the conditions \eqref{RWACondition} depend on both the ratio $R$ and its inverse. 
However, there is always a window of values $R$ that achieves near-unit fidelity.

\section{Derivation of the effective master equation}\label{app:effective-master-equation}
	In the main text we used an effective master equation for the mechanical degrees of freedom only, where the cavity field was adiabatically eliminated. Here we give a derivation of the effective master equation for a generic system consisting of a damped cavity mode coupled to an arbitrary function of the mechanical operators. The master equation describing the dynamics of joint density operator is
	\begin{equation}\label{eqn:full-master-eqn}
		\dot{\hat {\varrho}}=-i[\hat H,\hat\varrho]+\kappa\left(\hat a\hat \varrho \hat a^\dagger-\frac12 \hat a^\dagger\hat a\hat \varrho-\frac12\hat \varrho \hat a^\dagger \hat a\right)\ ,
	\end{equation}
	where $\hat H=G\left(\hat a^\dagger \hat f+\hat a \hat f^\dagger\right)$, $\hat a$ is the cavity mode annihilation operator, $\kappa$ is the cavity damping rate, and 
	$\hat f$ a function of $\hat b$ and $\hat b^\dagger$.
	We assume that there are two time scales in the system: a fast dynamics for the cavity and a slow one for the mechanical oscillator; this assumption translates into 
	$G\ll\kappa$. To eliminate the cavity variables we use the recipe for adiabatic elimination developed in \cite{azouit2016adiabatic}:
The master equation (\ref{eqn:full-master-eqn}) can be put in the form
	\begin{equation}\label{eqn:master-eqn-L0-L1}
		\dot{\hat {\varrho}}=\mathcal{L}_0\hat \varrho+G\mathcal{L}_1\hat\varrho\ ,
	\end{equation}
	where $\mathcal{L}_0\hat \varrho\equiv\kappa\left(\hat a\hat\varrho \hat a^\dagger-\frac12 \hat a^\dagger \hat a\hat \varrho-\frac12\hat \varrho \hat a^\dagger \hat a\right)$ and $\mathcal{L}_1\hat\varrho\equiv-i[\hat H_1,\hat\varrho]$, with $\hat H_1\equiv \hat H/G$. Then we treat the second term of Eq.~(\ref{eqn:master-eqn-L0-L1}) as a perturbation since we assumed that $G\ll\kappa$. We write the effective master equation for the mechanical oscillator in the form
	\begin{equation}
		\dot{\hat{\varrho}}_b=\mathcal{L}_b\hat \varrho_b\ ,
	\end{equation}
	where $\hat \varrho_b$ is the density operator describing the mechanical state, and $\mathcal{L}_b$ is a Lindbladian. The latter is expressed as a power series in the perturbation parameter $G$ 
	\begin{equation}
		\mathcal{L}_b\hat \varrho_b=\sum_{n\ge 1} G^n\mathcal{L}_{b,n}\hat \varrho_b\ .
	\end{equation}
	Up to second order in perturbation theory, $\mathcal{L}_{b,1}$ and $\mathcal{L}_{b,2}$ are given by the expressions \cite{azouit2016adiabatic}
	\begin{eqnarray}
		\mathcal{L}_{b,1}\hat \varrho_b	&=&	-i[\hat H_b,\hat\varrho_b]\ ,\\
		\mathcal{L}_{b,2}\hat \varrho_b	&=&	\sum_\ell\left(\hat B_\ell\hat\varrho_b \hat B_\ell^\dagger-\frac12\left\{\hat B_\ell^\dagger \hat B_\ell,\hat\varrho_b\right\}\right)\ ,
	\end{eqnarray}
	where $\hat H_b=\hat S^\dagger \hat H_1 \hat S$ and $\hat B_\ell=2 \hat S^\dagger \hat M_\ell \hat L\left(\hat L^\dagger \hat L\right)^{-1} \hat H_1 \hat S$ with $\hat L\equiv\sqrt{\kappa}\hat a$ and $\hat S$ and $\hat M_\ell$ are the operators defined as follows: in the absence of perturbation ($G=0$) the system evolves towards the steady state $\ket{0}_a \bra{0}\otimes \Tr{a}{\hat \varrho(0)}$. The set of all steady states (when the initial state $\hat \varrho(0)$ varies) has the support $\ket{0}_a\otimes\ket{\ell}_b$ ($\ell=0,1,\ldots$). The operator $\hat S$ is defined as $\hat S=\sum_\ell\left(\ket{0}_a\otimes\ket{\ell}_b\right){_b\bra{\ell}}$ and $\hat M_\ell$ are obtained from the relation $\ket{0}_a \bra{0}\otimes \Tr{a}{\hat \varrho(0)}=\sum_\ell \hat M_\ell\hat\varrho(0)\hat M_\ell^\dagger$ with the condition $\sum_\ell \hat M_\ell^\dagger \hat M_\ell=\mathds{1}$ (the identity operator in the Hilbert space of the system). It is straight forward to obtain $\hat M_\ell=\ket{0}_a\bra{\ell}\otimes\mathds{1}_b$ ($\mathds{1}_b$ denotes the identity operator in the Hilbert space of the mechanical oscillator). With these expressions we find $\hat H_b=0$ and $\hat B_\ell=\frac{2}{\sqrt{\kappa}}\ \delta_{\ell,0} \hat f$, so that  the effective master equation reads
	\begin{equation}
		\dot{\hat \varrho}_b=\frac{4G^2}{\kappa}\left(\hat f \hat \varrho_b \hat f^\dagger-\frac12\left\{\hat f^\dagger \hat f,\hat \varrho_b\right\}\right)\ .
	\end{equation}
This is the reduced master equation Eq.~\eqref{Adiabatic} upon the identification $G\equiv G_1$, $\hat a\equiv \hat d$, $\hat \varrho_b\equiv \hat \varrho^{(m)}$ and $\hat f$ as in Eq.~\eqref{Modef}.

\par
\section{Implementation}\label{app:Implementation}
\begin{figure}[t!] 
\includegraphics[width=.7\columnwidth]{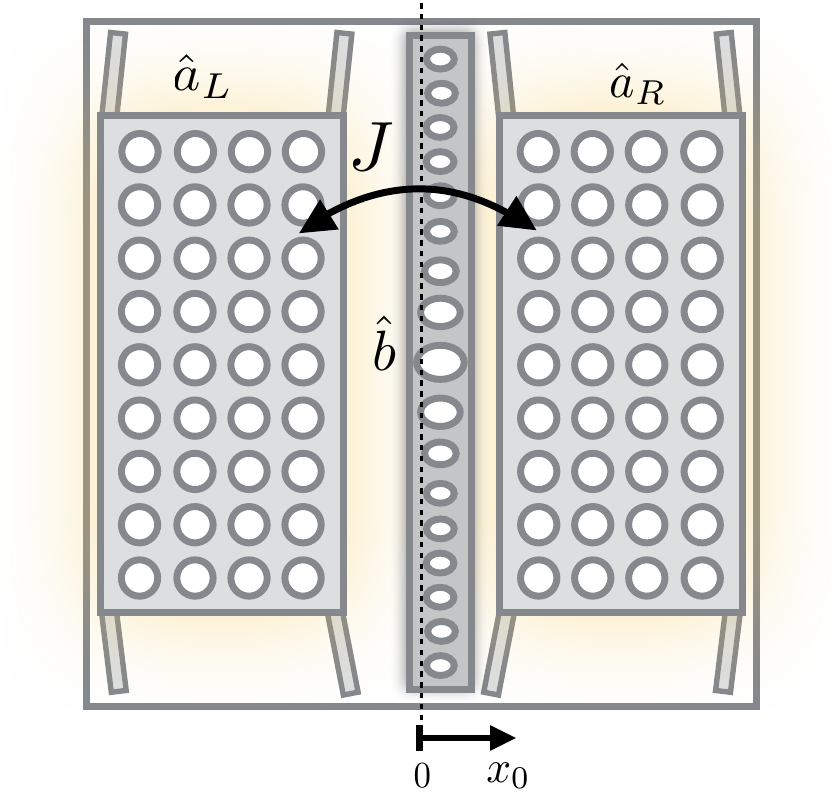}
\caption{Sketch of an optomechanical crystal implementation of a tunable linear-and-quadratic coupling. Due to photons hopping through the central nano-beam (mode $\hat b$), localized photonic modes $\hat a_{L,R}$ hybridize into supermodes delocalized over the two photonic cavities. The supermodes are optomechanically coupled to the nano-beam displacement and displacement squared. When the beam is equidistant from the two slabs ($x_0=0$), a purely quadratic optomechanical coupling is realized, while a controlled
offset of the nano-beam displacement $x_0\neq 0$, enables arbitrary linear-and-quadratic couplings.
\label{f:PlotOptoCrystal}}
\end{figure}

The system we consider for implementing our scheme consists of a pair two-dimensional photonic crystal cavities, obtained by patterning two thin 
silicon films, separated by a central suspended mechanical beam, also realized with a photonic crystal with a single row of holes (nano-beam). 
The two cavities  host  localized  degenerate optical modes $\hat a_L$ and $\hat a_R$ of the same frequency $\omega$ and 
are coupled at a rate $J$ via photon hopping across the central mechanical beam $\hat b$ of frequency $\omega_m$. 
A sketch of this multimode setup is given in Fig.~\ref{f:PlotOptoCrystal}.
Details about the experimental realization of such a device can be found in Ref.~\cite{NonLin5}, while  
Ref.~\cite{NonLin6} provides an in-depth study of the optical and acoustic modes accessible in this multimode structure and their 
optomechanical properties. 
The Hamiltonian of the three-mode optomechanical system is given by
		\begin{align}
			\hat H_{\mathrm{tot}}&=\hat H_0+\hat H_{\mathrm{hop}}+\hat H_{\mathrm{int}} \, ,\\
			\hat H_0&=\omega (\hat a_L^\dagger\hat a_L+\hat a_R^\dagger\hat a_R) +\omega_m\hat b^\dagger \hat b\,,\\
			\hat H_{\mathrm{hop}}&=J(\hat a_L^\dagger\hat a_R+\hat a_R^\dagger\hat a_L)\,,\\
			\hat H_{\mathrm{int}}&=x_{\mathrm{zpf}}(\hat b+\hat b^\dagger)(g_L \hat a_L^\dagger\hat a_L+g_R\hat a_R^\dagger\hat a_R)\,.
		\end{align}
		Due to the tunneling, the localized optical  modes hybridize into supermodes. The Hamiltonian written in the supermode basis $\hat a_\pm=(\hat a_L\pm \hat a_R)/\sqrt2$ can be diagonalized by assuming a quasi-static approximation of the mechanical motion, resulting in eigenfrequencies  $\nobreak{\omega_\pm=\omega_\pm(\hat x)}$ that are given by~\cite{SQND2}
		\begin{equation}\label{Omegapm}
			\omega_\pm(\hat x)=\omega+g_\pm \hat x\pm\sqrt{J^2+g_{+-}^2\hat x^2} \, ,
		\end{equation}
		where  $\hat x=x_{\mathrm{zpf}}(\hat b+\hat b^\dagger)$ and
		\begin{equation}
			g_+=g_-=\frac{g_L+g_R}{2}\quad \mathrm{and} \quad g_{+-}=\frac{g_L-g_R}{2}
		\end{equation}
		are referred to as linear self-mode coupling and linear cross-mode coupling, respectively. 
For the geometry we consider one has $g_L= -g_R$, so that by expanding Eq.~\eqref{Omegapm} around the position equidistant from the two slabs ($x_0=0$) one is left with a purely quadratic interaction with enhanced optomechanical coupling  $g_{0}^{(2)}=\frac{g_{+-}^2}{2J} x_\mathrm{zpf}^2$. The enhancement follows from the fact that  $J$ can be made arbitrarily small. On the other hand, when the central beam position is not equidistant from the two crystal cavities, i.e., the two air slots are not of the same width, the expansion of the supermode frequency around  $x_0\neq0$ leads to both a linear and a quadratic term. The expressions read~\cite{NonLin6}
				\begin{align}
		g_\pm(x_0)\approx& \frac{g_L+g_R}{2}\pm\frac{g_L-g_R}{2}\frac{Z}{\sqrt{Z^2+1}}\,, \\
		g_{+-}(x_0)\approx& \frac{g_L-g_R}{2}\frac{1}{\sqrt{Z^2+1}}\,, 
		\end{align}
		where $Z=\frac{(g_L-g_R)}{2J}x_0$,
which entail single-photon optomechanical coupling of the form
\begin{align}
g_0^{(1)}&=g_\pm(x_0)\,x_\mathrm{zpf}\, , \label{g01} \\ 
g_0^{(2)}&=\frac{g_{+-}^2(x_0)}{2J}\frac{1}{(Z^2+1)^{3/2}}\,x_\mathrm{zpf}^2\,. \label{g02}
\end{align}
The separation of the slots with respect to the central beam can be fine-tuned via electrostatic actuation, which provides extremely refined control over the ratio $R=g_0^{(2)}/g_0^{(1)}$.
To estimate the single-photon couplings we consider the following values, taken from the finite-element simulation of Ref.~\cite{NonLin6}:
$J/2\pi=0.1$ GHz, $x_\mathrm{zpf}=10$ fm, application of a bias voltage of a few tenth of millivolts that guarantees
a displacement $x_0<0.1$ pm (Ref.~\cite{Tunability} reports a measured tunability of $0.05$ $\mathrm{nm/V}^2$ in a similar double-slotted 
photonic crystal cavity), $g_L=-g_R=100$ GHz/nm. Plugging these parameters in Eqs.~\eqref{g01},~\eqref{g02} yields $g_0^{(2)}\approx 5$ kHz and $g_0^{(1)}\approx 70$ kHz,
and thus $R\approx0.07$, for which the RWA is an excellent approximation for several values of $n$ considered in Fig.~\ref{f:PlotCR}.
In general, depending on the specific target state $\ket{\varphi_n}$ to be stabilized, the ratio $R$ needs to be tuned to the required value(s).
\par
The central beam hosts several acoustic modes, both flexural 
mechanical resonances and localized `breathing' modes, ranging from a few MHz to a few GHz. A large
cavity quality factor of $Q\approx 4\times 10^6$ at the telecom wavelength $\lambda=1550$ nm places mechanical modes of frequency 
$\omega_m > 300$ MHz deep in the resolved-sideband regime. For the specific deign of Ref.~\cite{NonLin6}, finite-element simulations 
give for such high-frequency modes $x_\mathrm{zpf}\approx 3$ fm, which however is not much smaller than the one we have assumed
for the estimate of the bare optomechanical couplings. 
Photonic crystal cavities allow for large intra-cavity photon capacities $n_c>10^4$, which gives the multiphoton optomechanical couplings 
$G_{1,2,3}$  in the 10 -- 100 MHz range. Assuming a mechanical quality factor $Q_m=10^6$, a mechanical frequency $\omega_m/2\pi=400$ MHz, a cavity decay rate $\kappa/2\pi=50$ MHz, intra-cavity photon number $n_c=10^4$ and $g_0^{(1)}=70$ kHZ gives a multiphoton cooperativity $\mathcal{C}\approx 10^4$.
This would allow for the stabilization of our target state with high fidelity also without initialization in the ground state (see Fig.~\ref{f:PlotFidelity}). 
To give a reference, at the dilution refrigerator temperature of 15 mK the thermal occupation of a mode $\omega_m/2\pi=400$ MHz ($1$ GHz) is $\bar n=0.39\;(0.04)$, which increases to
 $\bar n=52\;(20)$ at 1 K. 
Larger values of the cooperativity can be obtained by considering higher mechanical quality factors and/or higher couplings.

\par
\section{Derivation of the steady-state solution}\label{app:Solution}
\begin{figure}[t!] 
\includegraphics[width=1.\columnwidth]{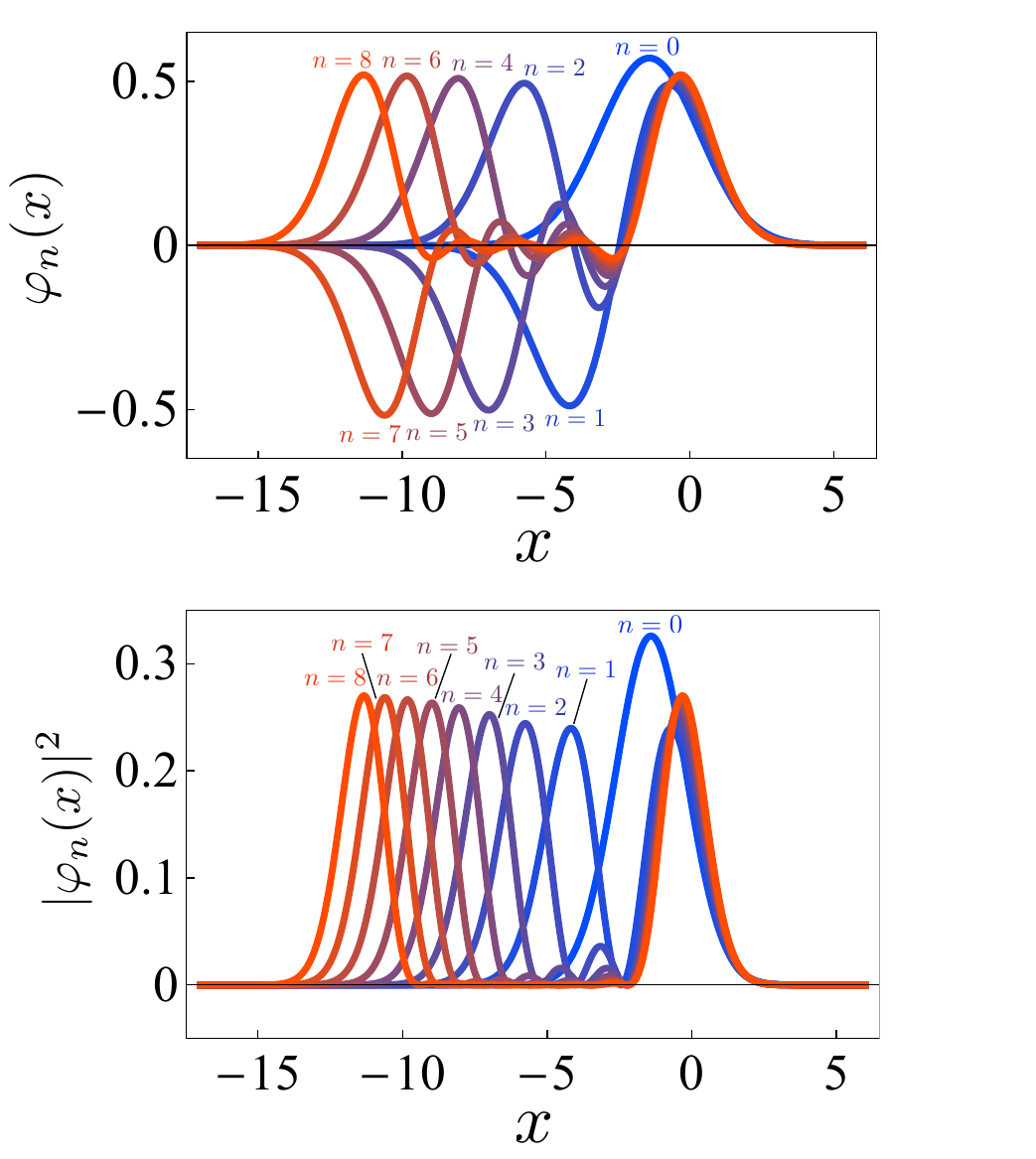}
\caption{Wave function $\varphi_n(x)$ (top) and corresponding probability density (bottom) 
from $n=0$ (blue) to $n=8$ (orange).
\label{f:PlotWaveFunc}}
\end{figure}
In this Section we derive the analytic expressions for the wave function Eq.~\eqref{Solution} and from that obtain the 
Fock state decomposition presented in Eq.~\eqref{NewState}. We also discuss how a finite accuracy in tuning the 
coefficients to the values prescribed by Eq.~\eqref{SChoice} affects the target state.
\subsection{Wave function}
The dark state condition Eq.~\eqref{DarkState} relative to the combination of mechanical creation and 
annihilation operators Eq.~\eqref{SFMode} can be equivalently expressed as the following differential equation for 
the system wave function $\varphi(x)=\langle x\ket{\varphi}$
\begin{align}\label{SDiffEq}
&\left(\tfrac{G_2}{2}-G_3\right) \varphi''(x) +
\left(\tfrac{G_1}{\sqrt{2}}-G_2x\right)\varphi'(x) \nonumber \\
&+\left[-\tfrac{G_2}{2}+\tfrac{ G_1}{\sqrt{2}}x+\left(\tfrac{G_2}{2}+G_3\right)x^2\right]\varphi(x) =0 \,.
\end{align}
This is a second order linear, homogeneous equation, whose  only square integrable solution (for suitable values of 
the coefficients $G_{1,2,3}$)  comes in the form of  a Hermite function,~i.e.~a Hermite polynomial times a Gaussian 
function. The explicit expression is rather involved, and hence not reported. We can simplify it by demanding that the 
order of the Hermite polynomial, which is expressed as a combination of  $G_{1,2,3}$, reduces to a non-negative 
integer value $n\in \mathbb{N}_0$.  This constraint can be expressed, e.g. as $G_2=G_2(n,G_1,G_3)$.  Moreover, 
upon direct inspection of the solution one can see that the expression greatly simplifies by choosing $G_3$ and 
$G_2$ equal and {\it opposite}. This choice fixes the form of the coefficients, whose magnitude is given by 
$\vert G_2\vert=\vert G_3\vert=\frac{G_1}{2\sqrt{2n+1}}$ and in the following we consider the case $G_3>0$, as 
shown in Eq.~\eqref{SChoice}.  As a result, the wave function acquires a universal character, depending only on 
the parameter $n$,  and takes the remarkably simple form
\begin{equation}\label{SSolution}
\varphi_n(x)=\mathcal{N}_n e^{-\frac{X_n^2}{4}} H_n(X_n)\, , 
\end{equation}
where we introduced  $X_n=\sqrt{\frac{2}{3}}\left(x+\sqrt{4n+2}\right)$ and  $\nobreak{\mathcal{N}_n=
(3\pi)^{-\frac14}\sqrt{\frac{n!}{(2n)! _2F_1(-n,-n;-n+\frac12;-\frac12)}}}$ is the normalization constant,   
$_2F_1(a,b;c;z)$ being the Gaussian hypergeometric function of argument $z$. We stress that  values of the ratio 
between the quadratic terms $G_3$ and $G_2$ different from that in Eq.~\eqref{SChoice} also lead to legitimate 
wave functions (for some the solution of Eq.~\eqref{DarkState} no longer describes a pure state), whose properties 
however may be very different from those of $\varphi_n(x)$ and whose nonclassical features are generally suppressed.
\par
Fig.~\ref{f:PlotWaveFunc} shows plots of the wave function Eq.~\eqref{SSolution} for different values of $n$ (left panel), 
together with the corresponding probability density function (right panel). We notice how the wave functions relative to 
an even/odd integer $n$ have distinct parity, as for the case of a simple harmonic oscillator. However, compared to the latter, 
the central oscillations of $\varphi_n(x)$ are progressively suppressed for increasing values of $n$ and at the same time the 
probability density develops a distinct bimodal character; this feature witnesses the transition to a  Schr\"{o}dinger
cat-like state for increasing $n$.
\par
\begin{figure*}[t!] 
\includegraphics[width=\linewidth]{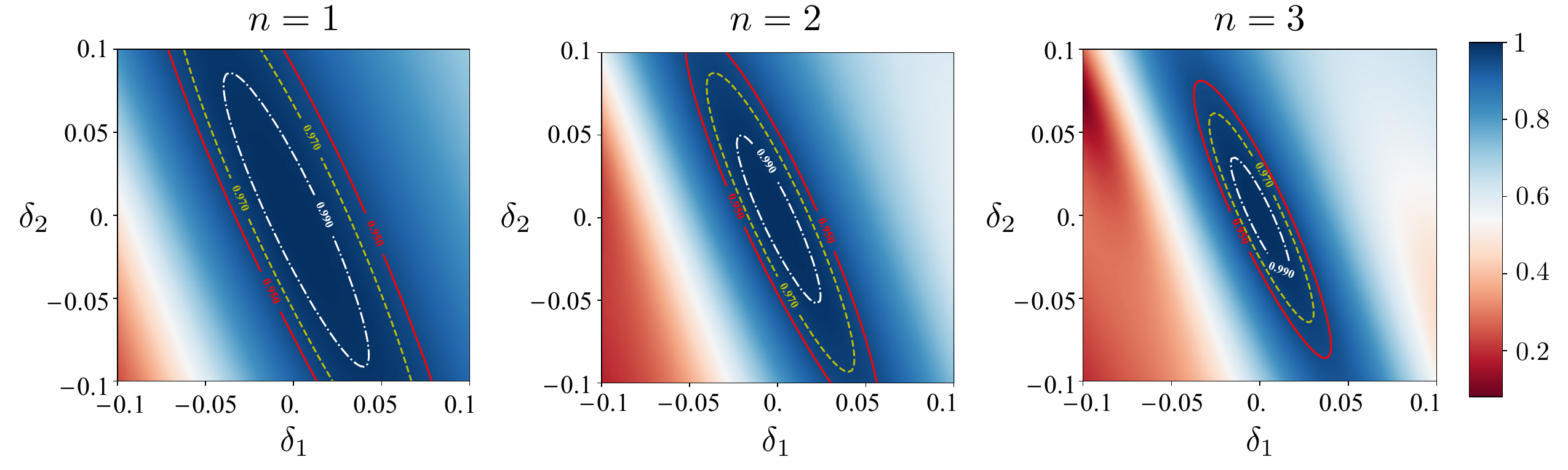}
\caption{Fidelity between the target state Eq.~\eqref{NewState} and the steady state obtained with perturbed 
couplings Eq.~\eqref{SPerturbed}. The fidelity is plotted against relative errors $\delta_1$ (horizontal axis) and $\delta_2$ 
(vertical axis) for $n = 1$ (left) $n = 2$ (centre) and $n = 3$ (right).
\label{f:PlotDeviations}}
\end{figure*}

Unlike the  quantum harmonic oscillator, where integer values labelling the solutions follow from the quantization of energy 
levels and the wave functions form an orthonormal set, in our case there are no fundamental mechanisms forbidding 
non-integer values -- these being determined by the choice of the drives -- and different $\varphi_n(x)$ are not orthogonal. 
The similarities between the two wave functions are due to the fact that the dark state condition Eq.~\eqref{SDiffEq} 
resembles the  Hermite differential equation encountered in the stationary Schr\"{o}dinger equation for a harmonic potential. 
However, since the ratio between the coherent drives can only be tuned up to a finite precision, it is important to verify that 
the target state is well-behaved with respect to imperfections. We then proceed to include small deviations from the optimal 
couplings shown in  Eq.~\eqref{SChoice}
\begin{equation}\label{SPerturbed}
G_2=G_1\left(\frac{-1}{2\sqrt{2n+1}}+\delta_1\right) ,\, G_3=G_1\left(\frac{1}{2\sqrt{2n+1}}+\delta_2\right),
\end{equation}
and check the deviation of the steady state from the ideal one. Fig.~\ref{f:PlotDeviations} confirms that the state is robust with 
respect to imprecisions in the strength of the drives. 
\par

\subsection{Fock state representation}
We now derive the explicit decomposition of the state $\ket{\varphi_n}$ in the Fock basis. To achieve this goal, we start 
from the expression of the wave function Eq.~\eqref{SSolution} and exploit the rescaling property of the Hermite 
polynomials that, for any $\gamma\in\mathbb{R}$, is given by
\begin{equation}
H_n(\gamma x)=\sum_{j=0}^{\lfloor \frac{n}{2}\rfloor} \gamma^{n-2j}(\gamma^2-1)^j\binom{n}{2j}\frac{(2j)!}{j!}H_{n-2j}(x)\, .
\end{equation}
The wave function is thus rewritten as 
\begin{subequations}
\begin{align}
\varphi_n(x)&=\mathcal{N}_n e^{-\frac12 \left(\frac{X_n}{\sqrt2}\right)^2} H_n\left(\sqrt2\, \frac{X_n}{\sqrt2}\right)\, , \label{SManip1} \\
&=\mathcal{N}_n\sum_{j=0}^{\lfloor \frac{n}{2}\rfloor} 2^{\frac{n}{2}-j} \binom{n}{2j} \frac{(2j)!}{j!}   
e^{-\frac12 \left(\frac{X_n}{\sqrt2}\right)^2} H_{n-2j}\left(\frac{X_n}{\sqrt2}\right)\, , \label{SManip2}
\end{align}
\end{subequations}
where $\lfloor y\rfloor$ is the floor function of argument $y$. We hence see that the wave function Eq.~\eqref{SSolution} is in fact 
a superposition of $\lfloor \frac{n}{2}\rfloor+1$ harmonic oscillator wave functions of argument $X_n/\sqrt2$. Moreover, each of 
these is easily identified with the wave function of a squeezed displaced number state. Indeed, one finds
\begin{equation}\label{SDispSq}
\bra{x}\hat D(\zeta)\hat S(r)\ket{n}=\frac{1}{\pi^{\frac14}\sqrt{2^{n}n!e^r}}e^{-\frac12 \left(\frac{x+\sqrt2 \zeta}{e^r}\right)^2} 
H_n\left(\frac{x+\sqrt2 \zeta}{e^r}\right) ,
\end{equation}
where $\hat D(\zeta)$ and $\hat S(r)$ are displacement and squeezing transformations that, for real parameters  
reduce to $\nobreak{\hat D(\zeta)=e^{-i\sqrt2\zeta \hat p}}$ and $\hat S(r)=e^{-i\frac{r}{2}(\hat x\hat p+\hat p\hat x)}$.
Therefore, combining  Eqs.~\eqref{SManip2} and~\eqref{SDispSq} we can write 
\begin{align}\label{SSecondLast}
\varphi_n(x)&=\pi^{\frac14}\mathcal{N}_n\sum_{j=0}^{\lfloor \frac{n}{2}\rfloor} 2^{\frac{n}{2}-j} \binom{n}{2j} \frac{(2j)!}{j!} 
\sqrt{2^{n-2j}(n-2j)!e^r}\nonumber \\
&\times \bra{x}\hat D(\zeta_n)\hat S(r)\ket{n-2j}\, ,
\end{align}
where we set
\begin{equation}
r=\frac12 \ln3\,, \quad \text{and} \quad \zeta_n=-\sqrt{2n+1}\,.
\end{equation}
From Eq.~\eqref{SSecondLast} we can finally read the expression for the state in the Fock basis 
\begin{equation}
\ket{\varphi_n}=\mathcal{M}_n\hat D(\zeta_n)\hat S(r)\sum_{j=0}^{\lfloor \frac{n}{2}\rfloor}\frac{1}{2^{2j}j!\sqrt{(n-2j)!}}\ket{n-2j} \, ,
\end{equation}
where now the normalization factor reads $\mathcal{M}_n=\sqrt{\frac{n!}{_2F_1\left(\frac{1-n}{2},\frac{-n}{2};1;\frac14\right)}}$.
It is also clear that reversing the sign between $G_2$ and $G_3$ amounts to change of displacement direction.
\par
\section{Comparison with Schr\"{o}dinger cat states}\label{app:Comparison}
\begin{figure}[b!] 
\includegraphics[width=.9\columnwidth]{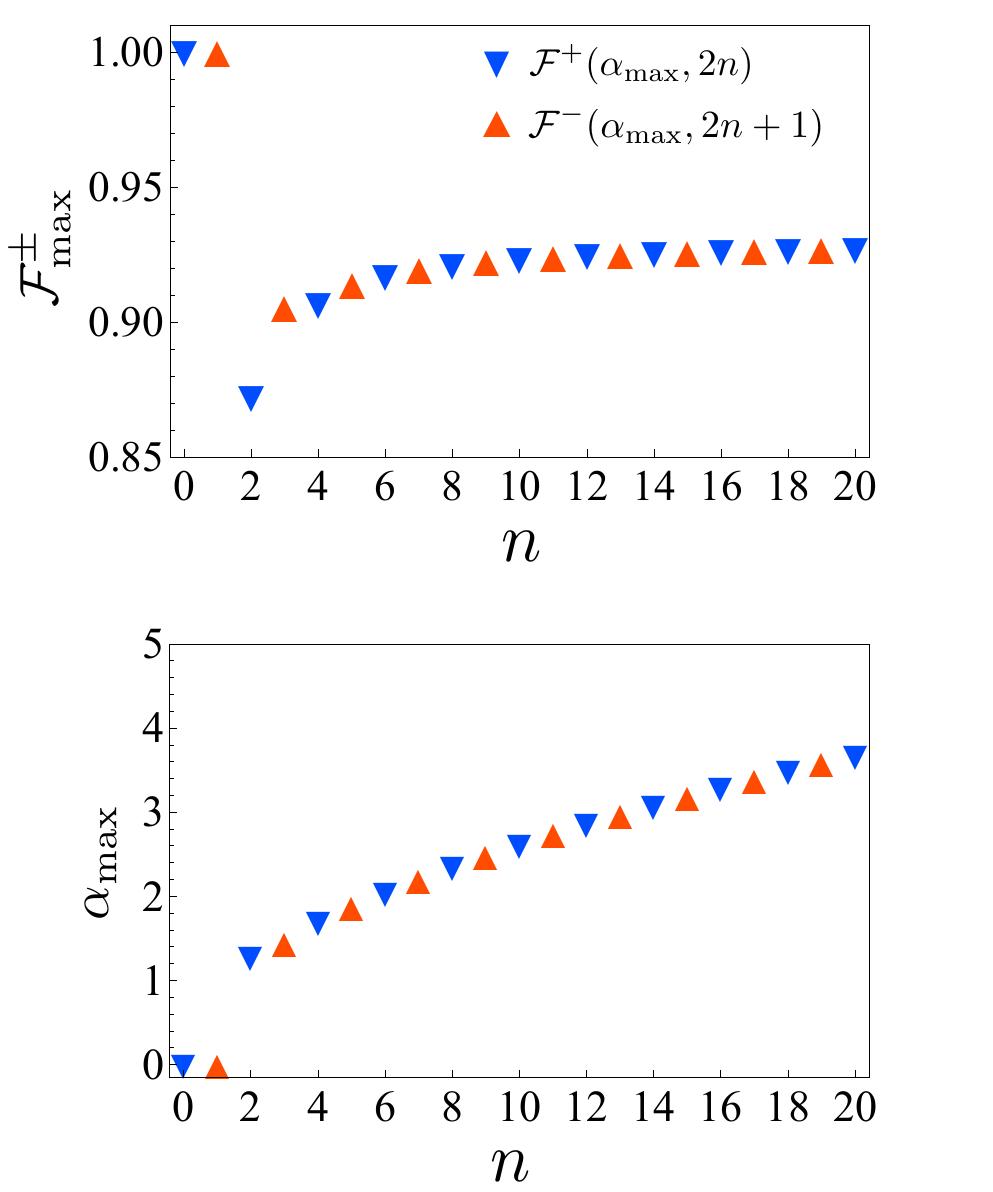}
\caption{
(Top) maximum fidelity between even (odd) Schr\"{o}dinger cat states and even (odd) finite superpositions for different 
integers $n$. (Bottom) values of the amplitude of the cat state yielding optimal fidelity for each $n$.  
\label{f:PlotCatFidelity}}
\end{figure}
We are now interested in comparing the target state of our protocol with a Schr\"{o}dinger cat state, which is a 
well-known benchmark for macroscopic quantum superposition states. We consider cat states of the following form 
\begin{equation}
\ket{\mathcal{C}_{\alpha}^\pm}=\mathcal{N}^\pm_{\alpha}(\ket{\alpha}\pm\ket{-\alpha}) \, ,
\end{equation}
where the normalization factor is given by $\mathcal{N}^\pm_{\alpha}=
\bigl[2\bigl(1\pm e^{-2\vert\alpha\vert^2}\bigr)\bigr]^{-\frac12}$ and the plus (minus) sign selects an even (odd) 
cat state, namely a superposition of only even (odd) number states. For a better comparison we also consider 
the target state $\ket{\varphi_n}$ without the squeezing and the displacement term, thus focusing on 
the finite superposition. The fidelity between the two states is computed as  $\mathcal{F}^\pm(\alpha,n)=\vert 
\bra{\mathcal{C}_{\alpha}^\pm}\widetilde{\varphi_n}\rangle\vert$, where  
\begin{equation}
\ket{\widetilde{\varphi_n}}
=\mathcal{M}_n\sum_{j=0}^{\lfloor \frac{n}{2}\rfloor}\frac{1}{2^{2j}j!\sqrt{(n-2j)!}}\ket{n-2j} \, .
\end{equation}
Given that both states have definite parity, the only nonzero overlaps  are between 
an even/odd cat state and an even/odd superposition of Fock states, and their expressions read
\begin{subequations}
\begin{align}
\mathcal{F}^+(\alpha,2n)&=\frac{\mathcal{M}_{2n}}{\sqrt{\cosh \vert\alpha\vert^2}}
\left\vert \sum_{j=0}^{n}\frac{(\alpha^*)^{2(n-j)}}{2^{2j}j!(2(n-j))!} \right\vert \, ,\\
\mathcal{F}^-(\alpha,2n+1)&=\frac{\mathcal{M}_{2n+1}}{\sqrt{\sinh \vert\alpha\vert^2}}
\left\vert \sum_{j=0}^{n}\frac{(\alpha^*)^{2(n-j)+1}}{2^{2j}j!(2(n-j)+1)!} \right\vert \, .
\end{align}
\end{subequations}
In Fig.~\ref{f:PlotCatFidelity} we show the maximum fidelity 
$\mathcal{F}^{\pm}_{\mathrm{max}}=\mathcal{F}^{\pm}(\alpha_{\mathrm{max}},n)$, optimized over 
$\alpha$, between an even (odd) cat state and even (odd) finite superposition of Fock states. The 
fidelity always lies within the range $\mathcal{F}^{\pm}_{\mathrm{max}} \approx 0.9-1$, providing 
further evidence that our state is indeed a macroscopic quantum superposition; larger values of $n$ 
correspond to larger superposition states, as also witnessed by the increasing amplitude of the ``closest'' 
cat state, shown on the right panel. However, by increasing $n$, the fidelity does not approach 1 and 
in fact saturates to a value $\mathcal{F}^{\pm}_{\mathrm{max}}\approx 0.92$, thus confirming that 
$\ket{\mathcal{C}_{\alpha}^\pm}$ and $\ket{\varphi_n}$ provide similar but always {\it distinct} instances 
of macroscopic superposition states.
\par
\section{Effects of the mechanical damping on the nonclassicality}\label{app:NonClassVolume}
\begin{figure*}[t!] 
\includegraphics[width=\linewidth]{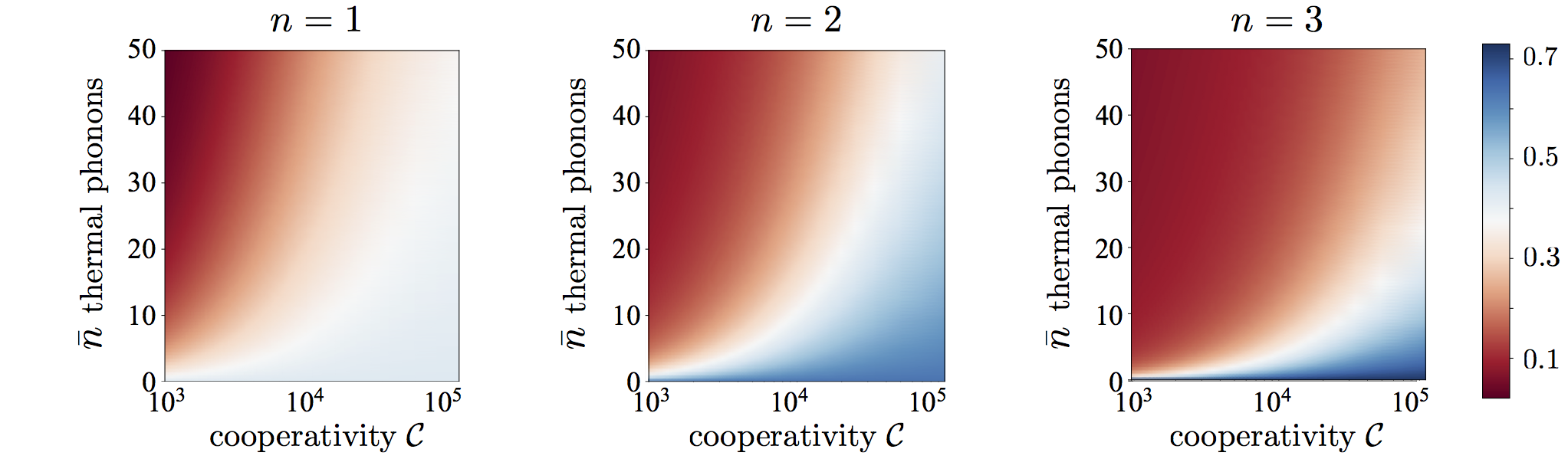}
\caption{Negative volume of the Wigner function [of the steady state of Eq.~\eqref{Adiabatic}] 
in the presence of mechanical damping  as a function of $\bar n$ and $\gamma$ (parametrized by the 
cooperativity $\mathcal{C}$), for $n = 1$ (left) $n = 2$ (centre) and $n = 3$ (right). $\nobreak{(G_1=0.05\kappa)}$.
\label{f:PlotNegVolume}}
\end{figure*}
We now address how the nonclassical features of the target state Eq.~\eqref{NewState} are affected 
by the presence of mechanical damping. To this aim, we consider the volume of the negative portion 
of the Wigner function, i.e.~$\nu^{(-)}=\int_{\mathbb{R}^2} \mathrm{d}x\mathrm{d}p W(x,p)^{(-)}$, 
where $W(x,p)^{(-)}=\frac12\left\{\vert W(x,p)\vert -W(x,p)\right\}$, which is known to provide an indicator 
of the nonclassicality of the state~\cite{SNegWig}. In the limit $\gamma\rightarrow0$ the Wigner function 
$\nobreak{W(x,p)=\frac{1}{\pi}\int\mathrm{d}x\, e^{-2ipy}\varphi_n(x+y)\varphi_n^*(x-y)}$ can be expressed 
analytically from Eq.~\eqref{SSolution}, although for the general state $\ket{\varphi_n}$ its form is quite 
cumbersome  and hence not reported. For $\gamma\neq0$ we numerically obtain  $\hat \varrho_{ss}^{(m)}$ 
as the solution of $\mathcal{D}_{f}\bigl[\hat \varrho_{ss}^{(m)}\bigr]=0$ and compute its Wigner function. 
Notice that by definition, the quantity $\nu^{(-)}$ vanishes for nonclassical yet {\it Gaussian} states such as 
a squeezed vacuum state.  A plot of $\nu^{(-)}$ as a function of $\bar n$ and $\gamma$ (parametrized by 
the cooperativity $\mathcal{C}$) is shown in Fig.~\ref{f:PlotNegVolume}, for different values of $n$.
As expected, the negative volume is suppressed by the presence of mechanical dissipation and reduction 
of $\nu^{(-)}$ is more pronounced for increasing $n$.  However, the steady state is nonclassical for a large 
range of values, even when it  no longer  has near-unit fidelity with the target pure state Eq.~\eqref{NewState} 
(cf.~Fig.~\ref{f:PlotFidelity}). In particular, the state is non-Gaussian for all the values shown. 
\end{appendix}



\begin{thebibliography}{99}
\bibitem{Note}
Since our focus is on the preparation of non-Gaussian states, we take to be `nonclassical' a 
state that exhibits negativity in the Wigner function. A more stringent notion of nonclassicality, 
based on the negativity/singularity of the Glauber-Sudarshan $\nobreak{P\text{-function}}$, 
can be considered, which however includes Gaussian states, e.g. squeezed vacuum.
\end{thebibliography}
\end{document}